\documentclass[twocolumn,showpacs,preprintnumbers,amsmath,amssymb]{revtex4}

\usepackage{graphicx}
\usepackage{dcolumn}
\usepackage{bm}

\begin{document}


\title{Fractal design for an efficient shell strut under gentle compressive loading}

\author{Robert S. Farr}
 \affiliation{Unilever R\&D, Olivier van Noortlaan 120, AT3133, Vlaardingen, The Netherlands}
 \email{robert.farr@unilever.com}

\date{\today}

\begin{abstract}
Because of Euler buckling, a simple strut of length $L$ and Young modulus $Y$ 
requires a volume of material proportional to $L^3 f^{1/2}$ in order to 
support a compressive force $F$, where $f=F/YL^2$ and $f\ll 1$. By taking 
into account both Euler and local buckling, we provide a hierarchical 
design for such a strut consisting of intersecting
curved shells, which requires a volume of material proportional to the much
smaller quantity $L^3 f\exp\left[2\sqrt{(\ln 3)(\ln f^{-1})}\right]$.
\end{abstract}

\pacs{46.25.-y, 46.70.De}

\maketitle

\section{Introduction\label{section1}}
Fractals \cite{Mandelbrot} occur ubiquitously in nature and appear to arise in 
many different ways; examples from the physical and biological sciences include
colloidal flocculation\cite{Lin,Kantor}, percolation 
phenomena\cite{Isichenko} and the structure of transport networks in 
organisms\cite{West}.
In the area of structural mechanics, it has been claimed that the fractal 
morphology of trabecular bone is responsible in part for its
mechanical efficiency \cite{Huiskes1}; while the complex suture patterns of
ammonites of the Jurassic and Cretaceous periods have been conjectured to 
give greater strength to their shells\cite{Jacobs,Batt}.

Recent theoretical work on a highly simplified model system consisting of a 
brittle pressure-bearing plate has also suggested that a fractal structure 
can be highly efficient when the loading conditions are very gentle and the 
material very brittle \cite{Farr}. 
It is therefore of interest to explore other circumstances under which
fractal design principles can lead to high mechanical efficiency, in case 
there is a general theorem underlying the optimal design of elastic structures 
under gentle compressive loading. To this end, we consider here
the buckling behaviour of compression members. 

Consider first the classical case of an Euler strut \cite{Euler} in 
the form of a solid, cylindrical column of radius $r_0$ and length $L$,
made from an isotropic, linear elastic material of Young modulus $Y$ and 
Poisson ratio $\nu$,
and subject to a compressive force $F$ at the freely hinged ends.

We define two non-dimensional parameters: $f\equiv F/(YL^2)$, which is the 
compressive force scaled by $YL^2$, and $v=\pi r_0^2/L^2$, which is the volume 
of material used, scaled by $L^3$. We are interested in the regime of 
gentle loading, by which we mean $f\ll 1$ and $v\ll 1$.

Because of Euler buckling, the strut can only withstand forces such 
that $F<\pi^2 YI_0 /L^2$, where $I_0=\pi r_0^4/4$ is the second moment 
of the cross sectional area about the neutral axis of the 
beam \cite{Euler, Timoshenko}. Therefore the (non-dimensionalised) volume of
material required to withstand a load represented by $f$ is given by
\begin{equation}\label{euler}
v(0)=2\pi^{-1/2}f^{1/2},
\end{equation} 
where we have neglected the material required to make the freely hinged 
couplings at the ends of the strut. For the purposes of this paper, we 
call the above solid strut a ``generation $G=0$'' structure,
and this is written as an argument for the volume 
variable $v$ in Eq.~(\ref{euler}).

We note in passing that the contrast in efficiency between compression 
members and tension members (for which $v\propto f$) is a persistent 
theme in structural engineering. The scaling of Eq.~(\ref{euler}) plus the 
cost of couplings mean that efficient structures tend to have few 
compression members and many long tension members; the paradigmatic
example being a tent \cite{Cox,Gordon}.

To define a $G=1$ structure, we choose a hollow cylindrical shell, which is 
also a classic problem in elasticity theory \cite{Timoshenko,Koiter}. We 
choose the length as always to be $L$, and we denote the radius 
by $r_{1,1}$ where the first index refers to 
the ``generation number'' $G=1$ and the second index will be explained 
when we describe structures of higher generation number. This $G=1$ structure 
consists of one cylinder parallel to the compression direction, and we
express this trivial fact by $n_{1,1}=1$.  
The thickness of the sheet of elastic material making up the curved 
surface of the cylindrical shell is denoted by $s_{1,1}$, which 
specifically represents the volume of material required to make unit 
area of the curved surface.  We call this quantity the ``material thickness'' 
of the curved sheet. We also introduce 
an ``effective elastic thickness'' $t_{1,1}$ for the curved sheet. 
For the generation 1 structure, the curved sheet is simple in topology 
and uniform in thickness, and so $s_{1,1}\equiv t_{1,1}$.  
Lastly we have an effective Young modulus $Y_{1,1}$ and Poisson 
ratio $\nu_{1,1}$ for the sheet. For a $G=1$ structure $Y_{1,1}\equiv Y$ 
and $\nu_{1,1}\equiv\nu$. In all of these expressions, the first index 
refers to the generation number, and the second index will take values 
from $1$ up to the generation number of the structure, as will be 
explained in section \ref{generation2}.

Provided that $t_{1,1}\ll r_{1,1}\ll L$, the column now has a second 
moment of cross sectional area about the neutral axis given 
by $I_1=\pi s_{1,1}r_{1,1}^3$ and the volume of material used to 
construct it is given by $v=2\pi r_{1,1}s_{1,1}/L^2$. The requirement that
Euler buckling not occur then imposes the 
constraint $F<\pi^2 Y_{1,1}I_1 /L^2$ or
\begin{equation}
v>2f^{1/3}\left(t_{1,1}/L\right)^{2/3}.
\end{equation}

In contrast to the solid column, there is now the possibility of local buckling.
This happens when \cite{Koiter,Timoshenko}
\begin{equation}\label{koiter}
F=\frac{2\pi Y_{1,1}t_{1,1}^2}{\sqrt{3(1-\nu_{1,1}^2)}},
\end{equation}
which provides a second constraint on the structure required to 
support a force $F$.

The generation $1$ structure with the highest mechanical efficiency (smallest 
value of $v$ for a given $f$) is therefore specified by:
\begin{equation}\label{v1}
v(1)=2\left[\frac{3(1-\nu^2)}{4\pi^2}\right]^{1/6}f^{2/3}
\end{equation}
with
\begin{equation}
\frac{r_{1,1}}{L}=\frac{1}{\pi}\left[\frac{3(1-\nu^2)}{4\pi^2}\right]^{-1/12}f^{1/6},
\end{equation}
\begin{equation}
\frac{t_{1,1}}{L}=\left[\frac{3(1-\nu^2)}{4\pi^2}\right]^{1/4}f^{1/2}
\end{equation}
and therefore $t_{1,1}\ll r_{1,1}\ll L$ as assumed above.

Eq.~(\ref{v1}) represents a considerable gain in efficiency over the solid 
column of Eq.~(\ref{euler}), but does not rival the efficiency ($v\propto f$) 
which can be obtained for tensile loading.

We note that both here and in all subsequent sections, we have taken a 
conservative approach in calculating $v$, by assuming that the 
structure fails when the first buckling bifurcation is encountered. 
Engineering structures (especially shells) will often support considerably
higher loads in the post-buckling regime before catastrophic 
failure \cite{Timoshenko}. Although such complexities are certainly of 
practical importance, we choose to ignore them in this investigation 
and try, where possible, to obtain estimates which are upper bounds for $v$.

\section{A composite plate\label{plate}}
When designing a plate or shell which may buckle, it is standard 
engineering practice to introduce stiffening plates, longitudinal 
stringers, bulkheads, or similar devices in order to stiffen the 
structure and/or suppress buckling modes \cite{Timoshenko,Gordon}.

In this paper we take a similar approach, but re-design the structure in 
a systematic and hierarchical manner, which can be iterated in the 
limit $f\rightarrow 0$. We do not presume to do this in the most 
efficient manner (we are almost certainly over-engineering the 
protection against many of the buckling modes we wish to avoid) but 
nevertheless the design we describe allows us to systematically change 
the scaling of $v$ with $f$ and therefore to approach more closely the 
scaling which can be achieved for a rod under tension;
achieving ultimately $v\propto f\exp\left[2\sqrt{(\ln 3)(\ln f^{-1})}\right]$. 
In the limit $f\rightarrow 0$, this is smaller than
any scaling of the form $v\propto f^{\beta}$ with $\beta<1$.

To proceed in this direction, consider first of all a simple thin, flat plate 
of uniform thickness $\tilde{t}$, lying in the $x-y$ plane and made out of an 
isotropic elastic material of Young modulus $\tilde{Y}$ and Poisson 
ratio $\tilde{\nu}$. Suppose furthermore that this plate may be deformed by 
applied stresses, leading to stretching or shearing of the middle 
plane\cite{Timoshenko2}, and also
possibly to out-of-plane deflections which may be large compared to the 
plate thickness. An appropriate approximation to describe the behaviour of 
the plate is due to von Karman \cite{vonKarman}. In this theory, the 
equilibrium behaviour (given suitable boundary conditions) can be obtained
by minimising an energy functional. The elastic part of the energy (as
opposed to that from external forces) can be written as the
sum of two terms \cite{Lobkovsky}: the energy $U_S$ associated with stretching
of the middle plane of the plate, and a bending energy $U_B$.

If the 2-dimensional strain tensor for the middle plane of the plate
is given by ${\rm e}(x,y)$, then the stretching energy 
will be given by \cite{Timoshenko,Lobkovsky,Farr}
\begin{equation}\label{US}
U_{S}=\frac{\tilde{Y}\tilde{t}}{2(1-\tilde{\nu}^2)}\int dxdy\left\{
\tilde{\nu}\left[ {\rm Tr}({\rm e})\right]^2+(1-\tilde{\nu}){\rm Tr}({\rm e}^2)\right\}.
\end{equation}

Furthermore, if the plate is bent out-of-plane in the $z$-direction, by an 
amount $w(x,y)$, then the bending energy stored will be given 
by \cite{Lobkovsky}
\begin{equation}\label{UB}
U_{B}=\frac{\tilde{Y}\tilde{t}^3}{24(1-\tilde{\nu}^2)}\int dxdy\left\{
\left[ {\rm Tr}(H)\right]^2-2(1-\tilde{\nu}){\rm det}(H)\right\},
\end{equation}
where $H$ is the Hessian matrix
\begin{equation}
H(x,y)=\left(
\begin{array}{cc}
\frac{\partial^2 w}{\partial x^2} & \frac{\partial^2 w}{\partial x\partial y} \\
\frac{\partial^2 w}{\partial x\partial y} & \frac{\partial^2 w}{\partial y^2} 
\end{array}\right).
\end{equation}

\begin{figure}
\includegraphics[width=2.5in]{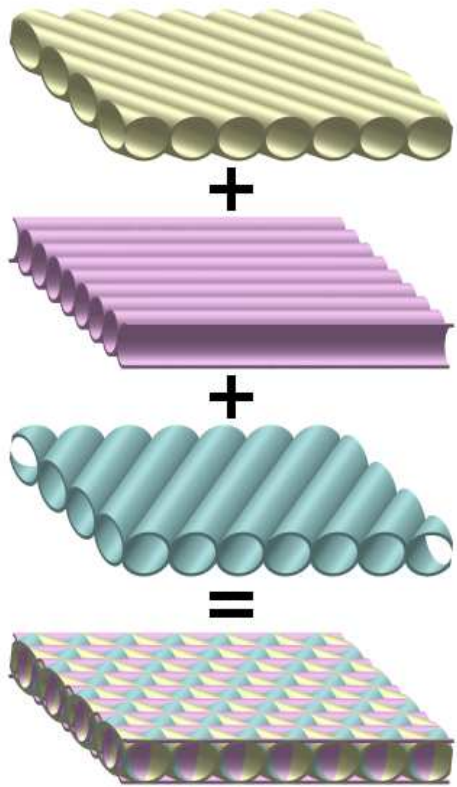}
\caption{\label{plate3}
Bottom image is part of a composite plate, which is constructed by merging the 
three substructures shown in the top three images of the figure. Each 
substructure is an infinite set of parallel right circular cylinders which are 
joined along their lines of osculation. The three sub-structures are 
identical save for being rotated by $\pm 2\pi/3$ radians relative to one 
another about an axis perpendicular to the plane. Each right circular 
cylinder has a 
radius $r$, a wall thickness $t\ll r$ and is infinitely long.}
\end{figure}

Now consider the composite plate shown in the bottom part of Fig.~\ref{plate3}. 
This structure is built from three intersecting sub-structures (as shown 
separately in the top three parts of Fig.~\ref{plate3}), each identical, save 
for being rotated $\pm 2\pi/3$ radians relative to one another. Where the
sub-structures pass through one another, we imagine them being joined or 
welded along their curves of intersection. Each sub-structure consists of 
an infinite set of parallel hollow right circular cylinders, with their 
axes in the $x-y$ plane, and placed so that each
touches two neighbours and is welded to each of its two neighbours along 
these lines of contact (which also lie in the $x-y$ plane). We specify all
these welds so we can be sure that on deformation at long length scales, the 
plate behaves
as a single entity, rather than separating into its constituent cylinders.

Each of the component cylinders has a radius $r$ and a wall 
thickness $t\ll r$. Because we have chosen the composite plate to have 
six-fold rotational symmetry about the $z$-axis,
then on long enough length scales the composite plate must behave elastically 
as though it is isotropic under rotations about the $z$-axis. This is because 
to leading order, the elastic stiffness of the plate under stretching 
and bending is represented by  
rank 4 tensors in two dimensions (which are contracted with 
two dimensional rank 2 deformation tensors to form the scalar energy).
These rank 4 elastic tensors may be invariant under rotation about the $z$-axis
(and so be consistent with any rotation group $C_n$), or may have 
one of the symmetry groups $C_2$ or $C_4$. However 
the composite plate we have described has the symmetry group $C_6$,
which is not consistent with 
$C_2$ or $C_4$.

Calculating the stretching and bending energies of the composite plate 
shown in the bottom panel of Fig.~\ref{plate3} is non-trivial even for the 
long-wavelength (much greater than $r$) deformations in which we are 
interested. Indeed, obtaining the correct numerical pre-factors would require 
a finite element calculation for the structure. Instead, and in order to 
proceed rigorously and in an analytical manner, we make an
approximation which under-estimates the stiffness (over-estimates
the compliance) of the structure, so that we would end up eventually 
with an upper bound on the minimum value of $v$ required for the final 
columns in the sections to follow.

The approximation we make is a ``ghost approximation'', in which we 
imagine that the cylinders composing the composite plate of Fig.~\ref{plate3} 
are no longer welded together, but are free to move past and indeed through 
one another; but all follow the imposed deformation field.

Under this approximation, consider what happens to one of the constituent 
cylinders when the composite plate is subjected to an in-plane stretching 
deformation, represented by a two-dimensional strain tensor ${\rm e}(x,y)$
with principal components $e_1$ and $e_2$. Let the cylinder be at an 
angle $\theta$ relative to the direction of the principal component $e_1$.
The cylinder will then be stretched parallel to its symmetry axis with a strain
\begin{equation}
\eta(\theta)=e_1\cos^2 \theta+e_2\sin^2 \theta.
\end{equation}
It may also rotate, but because of the ghost approximation, it experiences 
no resistance to this motion.

Adding up the stretching energies for the three sub-structures, we find a 
lower bound for the total stretching energy of the composite 
plate $U_{S,{\rm com}}$ given by
\begin{equation}\label{UScom}
U_{S,{\rm com}}\ge \frac{\pi Yt}{16}\int dxdy\left\{
3\left[ {\rm Tr}({\rm e})\right]^2+6{\rm Tr}({\rm e}^2)\right\}.
\end{equation}

Consider next a single cylinder when the composite plate is subjected to an 
out-of-plane bending deformation field $w(x,y)$, which is slowly varying 
in $x$ and $y$ (compared to the cylinder radius). If the cylinder is at an 
angle $\theta$ to the $x$-axis, then it will have an out-of-plane 
curvature given by
\begin{equation}
\kappa(\theta)=\frac{\partial^2 w}{\partial x^2}\cos^2\theta+
\frac{\partial^2 w}{\partial x\partial y}\sin 2\theta+
\frac{\partial^2 w}{\partial y^2}\sin^2\theta.
\end{equation}
From thin-beam theory\cite{Timoshenko} it will therefore have an 
elastic energy per unit length given by
\begin{equation}
u(\theta)=\frac{1}{2}Y\pi t r^3\left[\kappa(\theta)\right]^2.
\end{equation}
Adding up contributions from the three sub-structures, the bending 
energy $U_{B,{\rm com}}$ of the entire composite plate therefore 
has a lower bound given by
\begin{equation}\label{UBcom}
U_{B,{\rm com}}\ge\frac{3\pi Yr^2 t}{32}\int dxdy\left\{
3\left[ {\rm Tr}(H)\right]^2 -4 {\rm det}(H)\right\}.
\end{equation}

If we compare the lower bounds on elastic energy from Eqs.~(\ref{UScom}) 
and (\ref{UBcom}) with those for a uniform 
plate (Eqs.~(\ref{US}) and (\ref{UB}) ) then we see that for long wavelength 
deformations, the plate is at least as stiff as a uniform plate with effective 
thickness, Young modulus and Poisson ratio given by
\begin{eqnarray}
t_{\rm eff}=\sqrt{6} r, \label{teff} \\
Y_{\rm eff}=\frac{\pi}{\sqrt{6}}\left(\frac{t}{r}\right)Y, \label{Yeff} \\
\nu_{\rm eff}=1/3 \label{nueff}.
\end{eqnarray}

Furthermore, the composite plate uses an amount of material per unit area 
given by
\begin{equation}\label{s_recursion}
s_{\rm eff}=3\pi t.
\end{equation}

The results in Eqs.~(\ref{teff}), (\ref{Yeff}) and (\ref{nueff}) provide 
us with the information required to calculate lower limits on the stresses 
required to produce buckling on length scales much larger than $r$. 
However, a composite plate of this kind can also fail through
local buckling by one or more of the constituent cylinders undergoing 
a local buckling instability. 

This can be dealt with analytically for an isolated cylinder, 
which is the result used above for local buckling of a generation $1$ 
structure (Eq.~(\ref{koiter})). However, without an extensive
finite element study, covering a range of parameters, it is much 
harder to provide a good lower bound on the in-plane compressional 
stress required to excite these modes in the composite plate. 
This is because the stresses in intersecting cylinders could potentially 
generate buckling, rather than having no effect (as in the ghost 
approximation) or suppressing these modes.

In what follows, we will make the simple, but not necessarily 
accurate approximation that provided the largest compressive principal 
stress component lies in the direction of the axes of one of the 
substructures in the composite plate, then we will get local buckling only
under the same circumstances as for an isolated cylinder. This is also a 
kind of ``ghost'' approximation, in that the role of the other parts of 
the substructure are ignored, but although it is plausibly a conservative 
approximation, it no longer provides a strict bound on $v$.

\section{The generation $2$ structure\label{generation2}}
\begin{figure}
\includegraphics[width=2.5in]{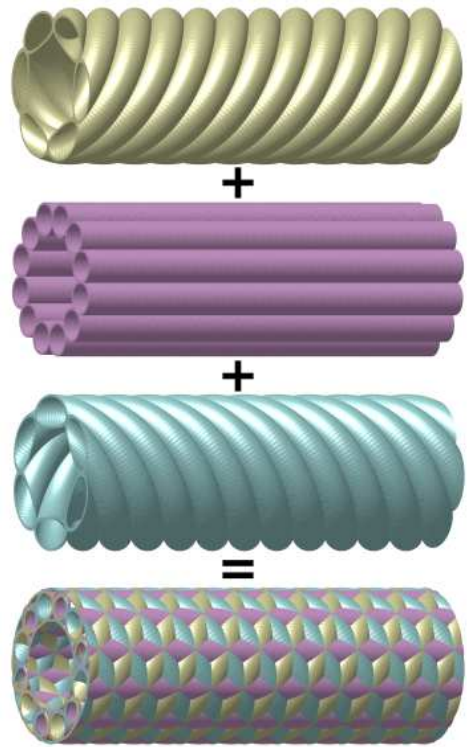}
\caption{\label{gen_2}
The bottom image is an example of a generation $2$ structure or column, which 
consists of three intersecting sub-structures that are shown in the top 
three images. Each sub-structure is composed of hollow cylindrical shells, 
with one sub-structure having the cylinders aligned parallel to the long 
axis of the entire column, and in the other two sub-structures the 
cylinders are wrapped in a helical arrangement (left handed for 
one sub-structure and right-handed for the other).}
\end{figure}

Imagine taking the curved cylindrical shell which forms the 
generation $1$ structure of section \ref{section1} and replacing the 
solid curved shell with a composite plate similar to that in
Fig.~\ref{plate3}, but curved to follow the original cylindrical 
surface. An example is shown in the bottom image of Fig.~\ref{gen_2}, 
with the top three images in the same figure showing the three 
substructures which are merged to form the final column; one of the
substructures has the axes of its constituent cylinders aligned with 
the long axis of the column itself. We refer to the result as
a generation $G=2$ structure.

\begin{figure}
\includegraphics[width=3in]{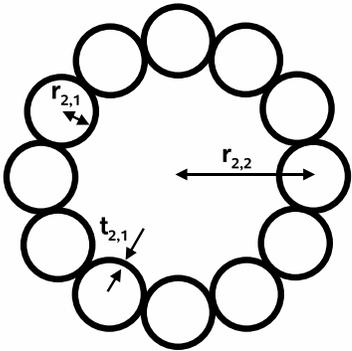}
\caption{\label{gen_2_cross_section}
Section through one of the sub-structures in Fig.~\ref{gen_2}, which 
has the component hollow cylinders parallel to the axis of the entire 
generation $2$ structure. The figure shows definitions
of various lengths required to specify a generation $2$ structure.}
\end{figure}

The length of the generation $2$ structure is taken to be $L$, and the 
radius of the entire column is $r_{2,2}$. At the largest scale, there 
is only ever one cylinder, and we represent this trivial fact 
by $n_{2,2}=1$. However, that substructure making up the composite shell 
which has cylinders aligned with the column length is composed of 
more than one cylinder, and we denote this number by $n_{2,1}$. For 
the structure in Fig.~\ref{gen_2}, we have $n_{2,1}=12$, as can be seen 
more clearly in the section through the relevant sub-structure
shown in Fig.~\ref{gen_2_cross_section}. 

The thickness of the thinnest shells, which make up the cylinders of the 
composite shell is $t_{2,1}$, and these have Young 
modulus $Y_{2,1}\equiv Y$ and Poisson ratio $\nu_{2,1}\equiv \nu$.
These shells form the cylinders of the composite sub-structures, which 
each have radius $r_{2,1}\gg t_{2,1}$. The resulting composite shell 
has an effective elastic thickness $t_{2,2}$, and effective Young modulus 
and Poisson ratio given by $Y_{2,2}$ and $\nu_{2,2}$.

The geometrical terms are illustrated in Fig.~\ref{gen_2_cross_section}, 
which for clarity shows only a cross-section through the sub-structure 
which has its cylinders aligned with the axis of the entire composite 
column. We note finally that in the case $r_{2,1}\ll r_{2,2}$ then
we can count the number of cylinders in this sub-structure using
\begin{equation}\label{n_recursion}
n_{2,1}=\pi r_{2,2}/r_{2,1}.
\end{equation}

Rather than analysing the efficiency of the $G=2$ structure here, we 
proceed directly to the general case in the following section.

\section{The generation $G$ structure}
To make the generation $3$ structure, we imagine replacing all the curved 
but solid shells in a generation $2$ structure (which are of uniform 
thickness $t_{2,1}$) with composite shells, as described for a 
flat plate in section \ref{plate} above.

The thinnest (solid) shells comprising this new structure have a 
thickness of $t_{3,1}$ and compose thin cylinders of 
radius $r_{3,1}$. These form the substructures of curved shells 
with effective elastic thickness $t_{3,2}$. These curved shells form cylinders 
of radius $r_{3,2}$ which compose the substructures of a new composite 
shell, which is of effective elastic thickness $t_{3,3}$, and from 
this final doubly-composite shell, a hollow cylinder of length $L$ and 
radius $r_{3,3}$ is formed, which is the final generation $3$ structure. 

By iterating this process, we end up with a generation $G$ structure. 
In general, $r_{G,m}$ is the radius of each of the (usually composite)
cylinders at hierarchical level $m$ of the generation $G$ structure:
$r_{G,1}$ is the radius of the smallest (non-composite) cylinders and
$r_{G,G}$ that of the multiply composite column itself. Similarly
$t_{G,m}$ is the effective elastic thickness of the curved (and usually 
composite) shells making up the cylinders at hierarchical level $m$ in the 
generation $G$ structure. 

We assume that
\begin{equation}\label{ineq}
t_{G,1}\ll r_{G,1}\lesssim t_{G,2}\ll r_{G,2}\lesssim \ldots \lesssim t_{G,G}\ll r_{G,G}\ll L
\end{equation}
and by definition we can say the following:
\begin{eqnarray}
t_{G,1}\equiv s_{G,1}, \label{t1}\\
Y_{G,1}\equiv Y, \label{Y1}\\
\nu_{G,1}\equiv \nu. \label{nu1}
\end{eqnarray}
The material thickness of the different composite shells are related to 
one another through Eq.~(\ref{s_recursion}) by 
\begin{equation}
s_{G,m}=3\pi s_{G,m-1} \label{sm}
\end{equation}
and therefore the total (non-dimensionalised) volume of material 
used is given by
\begin{equation}
v(G)=2\pi L r_{G,G}s_{G,G}/L^3. \label{vG}
\end{equation}
The effective elastic properties for $m\in\left\{ 2,\ldots ,G\right\}$ are 
related to one another in the ghost approximation (which gives an upper 
bound on $v$) by results analogous to Eqs. (\ref{teff}), (\ref{Yeff}) 
and (\ref{nueff}) of section \ref{plate}; namely
\begin{eqnarray}
t_{G,m}=\sqrt{6}r_{G,m-1}, \label{tm}\\
Y_{G,m}=\frac{\pi}{\sqrt{6}}\left(\frac{t_{G,m-1}}{r_{G,m-1}}\right)Y_{G,m-1}, \label{Ym}\\
\nu_{G,m}=1/3. \label{num}
\end{eqnarray}

At the largest length scale (i.e. the column itself), the structure is 
subject to Euler buckling and therefore the largest (non-dimensionalised) 
force it can support is subject to the constraint 
\begin{equation}\label{largest_euler}
f<\frac{1}{YL^2}\frac{\pi^2 I_G Y_{G,G}}{L^2},
\end{equation}
where the relevant second moment of the cross sectional area about the neutral axis is
given by
\begin{equation}
I_G=\pi t_{G,G} r_{G,G}^3.
\end{equation}

For the constraints due to localised buckling, we proceed as follows: at 
the largest length scale, there is one (multiply composite) cylinder 
aligned with the long axis of the column. This fact is captured by 
the equation $n_{G,G}=1$.

At the other length scales, we count the number of smaller cylinders 
in the substructures which are aligned with the long axis of the 
entire column in the following recursive manner, based 
on Eq.~(\ref{n_recursion}):
\begin{equation}
n_{G,m-1}=\pi\frac{r_{G,m}}{r_{G,m-1}}n_{G,m}, \label{nm}
\end{equation}
where $m\in\left\{2,\ldots ,G\right\}$.

Each of these cylinders at a level $m$ of the structure has a shell with an 
effective elastic thickness of $t_{G,m}$, effective Young modulus $Y_{G,m}$, 
effective Poisson ratio $\nu_{G,m}$ and supports a force no more than 
\begin{equation}\label{FGm}
F_{G,m}\equiv\frac{F}{n_{G,m}},
\end{equation}
which we obtain by neglecting the support provided by the other two 
helically arranged sub-structures at this level (and so on recursively).

As discussed in section \ref{plate} we make the crude approximation 
that the local buckling condition for a composite shell can be obtained 
from that for the isolated cylinders composing it. This again uses the 
ghost approximation, but in this case the approximation no longer 
provides a strict bound. The result is a sequence of conditions to avoid 
local buckling at each hierarchical level
in the structure, analogous to Eq.~(\ref{koiter}) and given by
\begin{equation}\label{koiter_hierarchy}
F_{G,m}<\frac{2\pi Y_{G,m} t_{G,m}^2}{\sqrt{3(1-\nu_{G,m}^2)}}.
\end{equation}

We now proceed to solve the recursion relations of 
Eqs.~(\ref{t1}-\ref{sm}), (\ref{tm}-\ref{num}) and (\ref{nm}), 
keeping first of all $t_{G,1}$ and $r_{G,m}$ as parameters for optimisation:
\begin{eqnarray}
s_{G,m}=(3\pi)^{m-1}t_{G,1} \label{sGm}\\
t_{G,m}=\left\{
\begin{array}{cc}
t_{G,1} & m=1 \\
\sqrt{6}r_{G,m-1} & 2\le m\le G
\end{array}
\right. \\
Y_{G,m}=\left\{
\begin{array}{cc}
Y & m=1 \\
\frac{\pi^{m-1}}{\sqrt{6}}
\left( \frac{t_{G,1}}{r_{G,m-1}}\right)Y & 2\le m\le G
\end{array}
\right. \\
\nu_{G,m}=\left\{
\begin{array}{cc}
\nu & m=1 \\
1/3 & 2\le m\le G
\end{array}
\right. \\
n_{G,m}=\pi^{G-m}\frac{r_{G,G}}{r_{G,m}}.\label{nGm}
\end{eqnarray}

The condition to be at the limit of Euler buckling (Eq.~(\ref{largest_euler}))
therefore becomes
\begin{equation}\label{wholeEuler}
f=\frac{\pi^{G+2}}{L^4} r_{G,G}^3 t_{G,1},
\end{equation}
the condition to be at the limit of local buckling at the smallest 
scale of the structure is (from Eqs.~(\ref{FGm}), (\ref{koiter_hierarchy})
and (\ref{nGm}))
\begin{equation}\label{wholeSmallLocal}
f=\frac{2\pi^G}{L^2\sqrt{3(1-\nu^2)}}\frac{r_{G,G}t_{G,1}^2}{r_{G,1}},
\end{equation}
and to be at the limit of local buckling at the other levels in the 
structure, gives (from Eqs.~(\ref{FGm}), (\ref{koiter_hierarchy}),
(\ref{nGm}) and (\ref{tm})) for $2\le m\le G$
\begin{equation}\label{wholeLocal}
f=\frac{3\pi^G}{L^2}
\frac{r_{G,m-1}r_{G,G}t_{G,1}}{r_{G,m}}.
\end{equation}

Eqs.~(\ref{vG}), (\ref{sGm}), (\ref{wholeEuler}), (\ref{wholeSmallLocal}) 
and (\ref{wholeLocal}) can be solved for $G>1$ to give finally
\begin{eqnarray}
v(G)=2^{(1+G)/(2+G)}3^{(2G^2-1)/[2(2+G)]}
\pi^{(G-2)/(2+G)} \nonumber \\
\times (1-\nu^2)^{1/[2(2+G)]} f^{1-[1/(G+2)]},\label{v_solution}
\end{eqnarray}
\begin{eqnarray}
\frac{t_{G,1}}{L}=2^{-3/[2(2+G)]}
3^{(9-6G)/[4(2+G)]} \pi^{-(2G^2-G+2)/[2(2+G)]} \nonumber \\
\times (1-\nu^2)^{3/[4(2+G)]} f^{(1+2G)/[2(2+G)]},
\end{eqnarray}
\begin{eqnarray}
\frac{r_{G,G}}{L}=2^{1/[2(2+G)]}
3^{(2G-3)/[4(2+G)]} \pi^{-(3G+2)/[2(2+G)]} \nonumber \\
\times (1-\nu^2)^{-1/[4(2+G)]} f^{1/[2(2+G)]},
\end{eqnarray}
while for $1\le m<G$
\begin{eqnarray}
\frac{r_{G,m}}{L}=2^{(2G-2m+1)/[2(2+G)]} 
3^{-(12G-14m+3)/[4(2+G)]} \nonumber \\
\times \pi^{-(2G^2-2Gm-G+4m+2)/[2(2+G)]} \nonumber \\
\times (1-\nu^2)^{-(2G-2m+1)/[4(2+G)]}
f^{(1-2m+2G)/[2(2+G)]}. \label{rGm_solution}
\end{eqnarray}

As a simple practical example, consider a strut of length $L=200{\rm m}$ 
which is required to support a force of $F=10{\rm kN}$, and which is 
made from a model material, similar to steel, 
with $Y=210{\rm GPa}$, $\nu=0.29$ and a density of $8000{\rm kgm}^{-3}$, so 
that $f=1.2\times 10^{-12}$. 

A cable supporting this force under tension would require a mass 
of $8{\rm kg}$ (assuming a 
yield stress for the material of $200{\rm MPa}$, and neglecting the mass of
couplings at the ends). The masses ($M$) of ``steel'' required for 
various structures described in this paper are shown in Table~\ref{steel}.

\begin{table}
\caption{\label{steel}
Example calculation for the mass $M$ if a structure required to 
support $F=10{\rm kN}$ over a distance of $L=200{\rm m}$ when the structure 
is made from a material similar to steel, with
$Y=210{\rm GPa}$, $\nu=0.29$ and density $\rho=8000{\rm kgm}^{-3}$.}
\begin{ruledtabular}
\begin{tabular}{cccccc}
$G$ & $M$ & $t_{G,1}$ & $r_{G,1}$ & $r_{G,2}$ & $r_{G,3}$ \\
\hline
0 & 79 tonnes & - & 12.5cm & - & - \\
1 & 941kg & 0.12mm & 81cm & - & - \\
2 & 319kg & 1.4$\mu$m & 1.2mm & 2.4m & - \\
3 & 261kg & 63nm & 14$\mu$m & 8.2mm & 4.6m \\
\end{tabular}
\end{ruledtabular}
\end{table}

\begin{figure}
\includegraphics[width=3in]{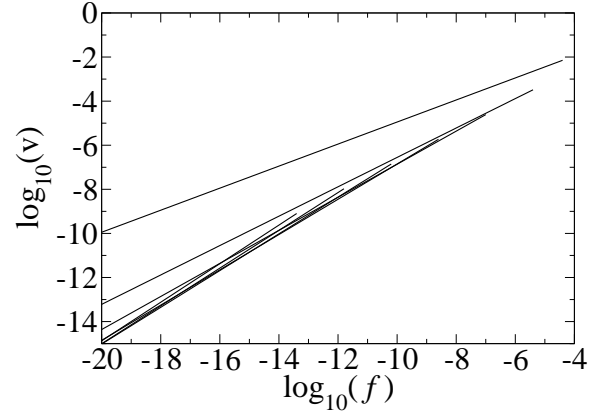}
\caption{\label{v_vs_f}
Plot of $\log_{10}(v)$ versus $\log_{10}(f)$ for the optimal structures with
$G=0,1,2,3,4,5,6$, using Eq.~(\ref{v_solution}) from the text and $\nu=0.29$. 
The curves are only drawn for the range of $f$ where $r_{G,G}/L\le 1/20$,
$r_{G,m}/r_{G,m+1}\le 1/20$ and $t_{G,1}/r_{G,1}\le 1/20$, which we take 
as an approximation to the condition of Eq.~(\ref{ineq}) in the text.}
\end{figure}

Lastly, we note that for a given value of $f$, several generations of 
structures may be compatible with the conditions of Eq.~(\ref{ineq}), 
as illustrated in Fig.~\ref{v_vs_f}. In the limit $f\rightarrow 0$, we 
can calculate the envelope of these curves in order to obtain the global 
optimally efficient structure within this class, through solving \cite{Courant}
\begin{eqnarray}
v=v(G) \label{envelope1} \\
\frac{\partial v(G)}{\partial G}=0, \label{envelope2}
\end{eqnarray}
where $v(G)$ is given by Eq.~(\ref{v_solution}).

In the limit of small $f$, we can expand the exponent of Eq.~(\ref{envelope1}) 
in powers of $1/G$ to obtain the assymptotic results
\begin{eqnarray}
v\sim \frac{2\pi f}{9} \exp\left[2\sqrt{(\ln 3)(\ln f^{-1})}\right], \\
G\sim \sqrt{\frac{\ln f^{-1}}{\ln 3}}.
\end{eqnarray}

\section{conclusions}
We have described compression members consisting of intersecting curved 
shells in a fractal or hierarchical arrangement which are highly 
mechanically efficient in the limit of light compressional loading.

Fractal designs for efficient plates under gentle pressure loading have 
recently been studied in Ref.~\cite{Farr}. In this work,
the fractal design arises from two competing tendencies in the structure:
Firstly there is a geometrical feature of the plate (narrow, tall spars)
which when developed to extremes can lead to very high mechanical 
efficiency. Secondly, there is a limit to how far this feature may 
be developed, which is ultimately a vulnerability to buckling. One spar
can however be used to provide partial support for another, and this leads
to the final hierarchical design.

The parallels with the problem of the present paper 
should be apparent, and so we suspect that fractal forms may be
a general property of optimal elastic structures under gentle and at
least partially compressive loading.

At this stage, we are not able to frame a precise mathematical conjecture, 
and we do not know how other parameters will figure in the analysis.
In previous work brittleness \cite{Farr} was important, and it seems highly 
probable
that in the present work, the amplitude of imperfections in either the
geometry or the uniformity of the loading could be crucial to determining the
mechanical efficiency \cite{Timoshenko,Timoshenko2}. 

We therefore hope that further examples, and perhaps ultimately theorems, 
will shed light on a problem whose geometrical
solutions promise to be useful and even beautiful.

\begin{acknowledgments}
The author wishes to thank E. G. Pelan for allowing me the occasional
freedom to take on problems which I have no hope of solving. 
The author also acknowledges the open source community for some
excellent software invaluable to this work. For example, Figures \ref{plate3} 
and \ref{gen_2} were prepared using the constructive solid geometry 
capabilities of `PovRay' (http://www.povray.org) and Fig.~\ref{v_vs_f} was 
prepared using `Grace' (http://plasma-gate.weizmann.ac.il/Grace/).
\end{acknowledgments}

\end{document}